\documentclass[a4paper,aps,prb,reprint,showpacs,twocolumns]{revtex4-1}
\usepackage{amsmath,amssymb,mathrsfs}
\usepackage[pdftex]{graphicx}
\usepackage{color}

\allowdisplaybreaks[4]

\begin{document}

\title{Finite-time quantum quenches in the XXZ Heisenberg chain}
\author{Benedikt Schoenauer}
\author{Dirk Schuricht}
\email{d.schuricht@uu.nl}
\affiliation{Institute for Theoretical Physics, Center for Extreme Matter and Emergent Phenomena, Utrecht University, Princetonplein 5, 3584 CE Utrecht, The Netherlands}
\date{\today}
\pagestyle{plain}

\begin{abstract}
We study the time evolution of the two-point correlation functions in the XXZ Heisenberg chain after a finite-time quantum quench in the anisotropy. We compare results from numerical simulations to ones obtained in the Luttinger model and find good agreement. We analyse the spreading of the correlations and the associated light-cone features. We observe a delay in the appearance of the light cone as compared to the sudden-quench setup, and link this delay to the properties of the quench protocol. 
\end{abstract}
\maketitle


\section{Introduction}
The past 20 years have seen tremendous progress in the experimental realisation and control of ultracold atomic and ionic systems in optical lattices.\cite{Lewenstein-07,Bloch-08,ReichelVuletic11,GrossBloch17} In particular, these systems allow the study of the almost unitary time evolution following the sudden change of the system parameters, a setup known as a quantum quench.\cite{CalabreseCardy06} This has in turn triggered a huge theoretical effort\cite{Polkovnikov-11,Eisert-15,GogolinEisert16} to understand the non-equilibrium dynamics of closed quantum systems.

Motivated by the ground-breaking experiment on the Quantum Newton's cradle,\cite{Kinoshita-06} one-dimensional many-particle models have been a particular focus of theoretical research. A paradigmatic model in this class of systems is provided by the spin-1/2 Heisenberg chain. For example, it fostered the investigation\cite{Wouters-14,Pozsgay-14} of the relaxation of quantum systems subject to an extensive number of conserved quantities and their relation to integrability and generalised Gibbs ensembles.\cite{Rigol-07} 

Another generic feature of the quench dynamics in quantum systems is the appearance of light cones in two-point functions, which is ultimately related to the boundedness of the spreading of informations.\cite{LiebRobinson72,Bravyi-06} More physically, the light cone can be linked to the propagation of entangled pairs of quasiparticles through the system,\cite{CalabreseCardy06,CalabreseCardy07} and thus is expected to appear quite generally. Indeed, the light-cone effect has not only been observed in numerical simulations on various one-dimensional systems,\cite{LauchliKollath08,Manmana-09,Barmettler-12,HaukeTagliacozzo13,Eisert-13,Bonnes-14,Carleo-14,dePaula-17,Despres-19} but also in experiments on ultracold gases.\cite{Cheneau-12,Langen-13,Richerme-14,Jurcevic-14}

As is well-known, the Heisenberg chain allows a description in terms of an effective field theory.\cite{Giamarchi04} This Luttinger-liquid theory is well suited to capture the behaviour of the system at low energies and large distances, which has also been established for the dynamics after a quantum quench.\cite{Barmettler-10,KRSM12,Pollmann-13,Coira-13} In particular, Collura, Calabrese and Essler\cite{Collura-15} provided a detailed analysis of the two-point spin correlation functions after a sudden quench. For the transverse correlation function they found a surprising level of agreement between numerical simulations and the predictions obtained in the Luttinger model, while the agreement for the longitudinal correlation function was much poorer. 

Unlike the majority of previous work, which focused mainly on sudden quantum quenches, we here consider more general quenches of finite duration $\tau$, during which the interaction in the system is modified. The finite quench time $\tau$ introduces an additional energy scale $\sim \tau^{-1}$, which is directly related to the quench protocol and therefore tunable. As such, it can be chosen to be of similar size as other energy scales of the system, including the band width, excitation gaps or relaxation rates. The interplay of the additional energy scale with the established ones may then bring about emergent quantum states beyond the ones accessible through sudden quench protocols.

For the above mentioned Luttinger model this finite-time quench protocol has been studied in several works.\cite{Dora-11,Pollmann-13,Bernier-14,DziarmagaTylutki11,PerfettoStefanucci11,Dora-12,Dora-13,BacsiDora13,Sachdeva-14,CS16,Porta-16,Chudzinski16,Bacsi-19} In particular, the light-cone spreading in two-point correlation functions was found\cite{Bernier-14,CS16} to be delayed as compared to the light cone after sudden quenches, with the delay being related to the length and form of the finite-time quench protocol. Comparisons between the Luttinger model predictions and numerical simulations for interacting microscopic
models have been limited so far,\cite{Pollmann-13,Bernier-14} with a detailed analysis of the light-cone spreading and aforementioned delay for the Heisenberg chain still missing.

The aim of the present manuscript is to provide this analysis, ie, we numerically calculate the quench dynamics in the XXZ Heisenberg chain after finite-time quenches and compare the results to the previously obtained analytical predictions from the Luttinger model. In particular, we focus on the transverse two-point function and the delay observed in the light-cone spreading. Overall we find good agreement between the predictions from the Luttinger model and our simulations, especially for the dependence of the delay on the quench duration. For the longitudinal correlation function the observed delay for moderate quench times is still well described by the Luttinger model, even thought the overall agreement between the simulations and field theory is much poorer.

This article is organised as follows: In Sec.~\ref{sec:methodsetup} we introduce the Heisenberg XXZ chain and define the quench protocols and correlation functions to be studied. In Sec.~\ref{sec:method} we briefly discuss details of our numerical simulations. Following this, in Sec.~\ref{sec:transverse} we present our main results, namely the discussion of the simulation results for the transverse two-point function, their comparison to the predictions from the Luttinger model, and an analysis of the observed delay of the light cone. This is followed in Sec.~\ref{sec:longitudinal} by a brief discussion of results for the longitudinal two-point functions, before we conclude in Sec.~\ref{sec:conclusions}.

\section{Model and quench protocol}\label{sec:methodsetup}
In this work we study the XXZ Heisenberg chain with the Hamiltonian
\begin{align}
  H(t)=J\sum_{i=1}^L\Bigl[S^x_{i} S^x_{i+1} + S^y_{i} S^y_{i+1} + \Delta(t) S^z_i S^z_{i+1}\Bigr],
  \label{eq:H}
\end{align}
where $S^a_i$, $a=x,y,z$, denote the spin-1/2 operators on lattice site $i$. For the implementation of the quench protocol we allow for a time-dependent anisotropy $\Delta(t)$. The equilibrium properties of the model \eqref{eq:H} are well known.\cite{Giamarchi04} For $\vert \Delta \vert < 1$ the system is quantum critical, with its low-energy properties described by Luttinger liquid theory, whereas for $\vert \Delta \vert > 1$ the ground state is ordered and the excitation energies are gapped. Throughout this work we restrict ourselves to the critical regime with $0\le\Delta<1$ and consider an $L$-site chain with periodic boundary conditions.

Below we study the time evolution of the system after a finite-time quench in the anisotropy, ie, over a finite time interval $\tau$ we change the parameter $\Delta(t)$ from its initial value $\Delta_0$ to its final value $\Delta$. The case $\tau=0$ corresponds to the well-studied sudden quantum quench. Here we consider the case of vanishing initial anisotropy, $\Delta_0=0$, corresponding to the non-interacting model when mapping \eqref{eq:H} to a model of spinless fermions via a Jordan--Wigner transformation. For times $t$ larger than the quench time $\tau$ the anisotropy (interaction) is kept at a constant value $\Delta$. More specifically, we consider two types of continuous protocols, first the linear quench 
\begin{align}
   \Delta(t) =\left\lbrace\begin{matrix}
    \Delta \frac{t}{\tau}, & t < \tau, \\[2mm]
    \Delta, & t \geq \tau,
  \end{matrix}\right. 
  \label{eq:linear}
\end{align}
and second the exponential quench
\begin{align}
  \Delta (t) = \left\lbrace \begin{matrix}
    \Delta \left[\exp(\log (2)t/\tau)-1\right], & t < \tau, \\[2mm]
    \Delta, & t\geq \tau. 
  \end{matrix}\right.
  \label{eq:exponential}
\end{align}
In the following, we present a numerical study of the dynamics during and following these quenches. As we will see, even thought the two protocols are very similar, the numerical data still show differences that can be traced back to the different time dependences.

As observables we consider the equal-time transverse and longitudinal correlation functions of the spin operators at a distance $\ell$, ie, 
\begin{equation}
\left\langle S^x_{i}(t) S^x_{i+\ell}(t)\right\rangle=\frac{1}{4}\Bigl[\left\langle S^{+}_{i}(t)S^{-}_{i+\ell}(t)\right\rangle
  +\left\langle S^{-}_{i}(t) S^{+}_{i+\ell}(t) \right\rangle\Bigr]
  \label{eq:transverse}
\end{equation}
and
\begin{equation}
\left\langle S^z_i (t) S^z_{i+\ell}(t) \right\rangle.
  \label{eq:longitudinal}
\end{equation}
We will compare our numerical results to analytical expressions for the correlation functions obtained in the Luttinger model\cite{CS16} as well as previous numerical simulations\cite{Collura-15} for sudden quenches. Before doing so, we briefly discuss details of our numerical simulations. 

\section{Numerical method}\label{sec:method}
For our numerical study, we employ a time-dependent density-matrix renormalisation group (DMRG) algorithm.\cite{Vidal04,WhiteFeiguin04,Daley-04,Schmitteckert04} Unlike many other time-dependent DMRG algorithms, we use a Krylov subspace method\cite{Schmitteckert04} to calculate the full matrix exponential of the Hamiltonian for the time evolution. This gives us the possibility to choose time steps of arbitrary size $\Delta t$ while the Hamiltonian is time-independent. In a first DMRG step, we have calculated the ground state of the initial system with $\Delta_0=0$ on a chain of length $L=80$. (We keep this system size throughout.) The ground-state energy density is found to be $E_0/(JL)=-0.318378704$, which constitutes a deviation from the exact value $E_0/(JL)=-1/\pi$ of $\Delta E_0 =7\cdot 10^{-6}$, indicating that finite-size effects are sufficiently small at this system size.

Starting from the ground state, we perform the first part of the time evolution over the quench duration $\tau$, where we choose a small step size of $\Delta t\approx J\tau/80$. For this particular time evolution, we use time-independent snapshots of the Hamiltonian $H(t_i < t < t_{i+1}) \simeq H(t_i)$. Additional calculations with smaller step sizes have been performed finding identical results, thus confirming that the chosen step size is sufficiently small to capture the details of time dependent anisotropy $\Delta(t)$. Alternatively, one could employ a Magnus expansion\cite{Blanes-09} of the time-dependent Hamiltonian, in which case larger step sizes would be feasible. 

For the second part of the time evolution after the quench, we use a larger constant step size $J\Delta t< 1$. An important aspect of a global quantum quench is the linear growth of the entanglement entropy in response to it. We therefore dynamically adjust the number of kept states per block $1400<N_{\text{cut}} \leq 14000$ in each DMRG step to ensure that the maximum amount of discarded entanglement entropy does not exceed $\delta S_{\text{max}} = 10^{-4}$. The large number of kept states is also necessitated by the criticality of the system, due to which the entanglement entropy grows as a logarithm of the system size $L$.

\section{Transverse two-point function}\label{sec:transverse}
\begin{figure}[t]
\begin{center}
 \includegraphics[width=0.45\textwidth]{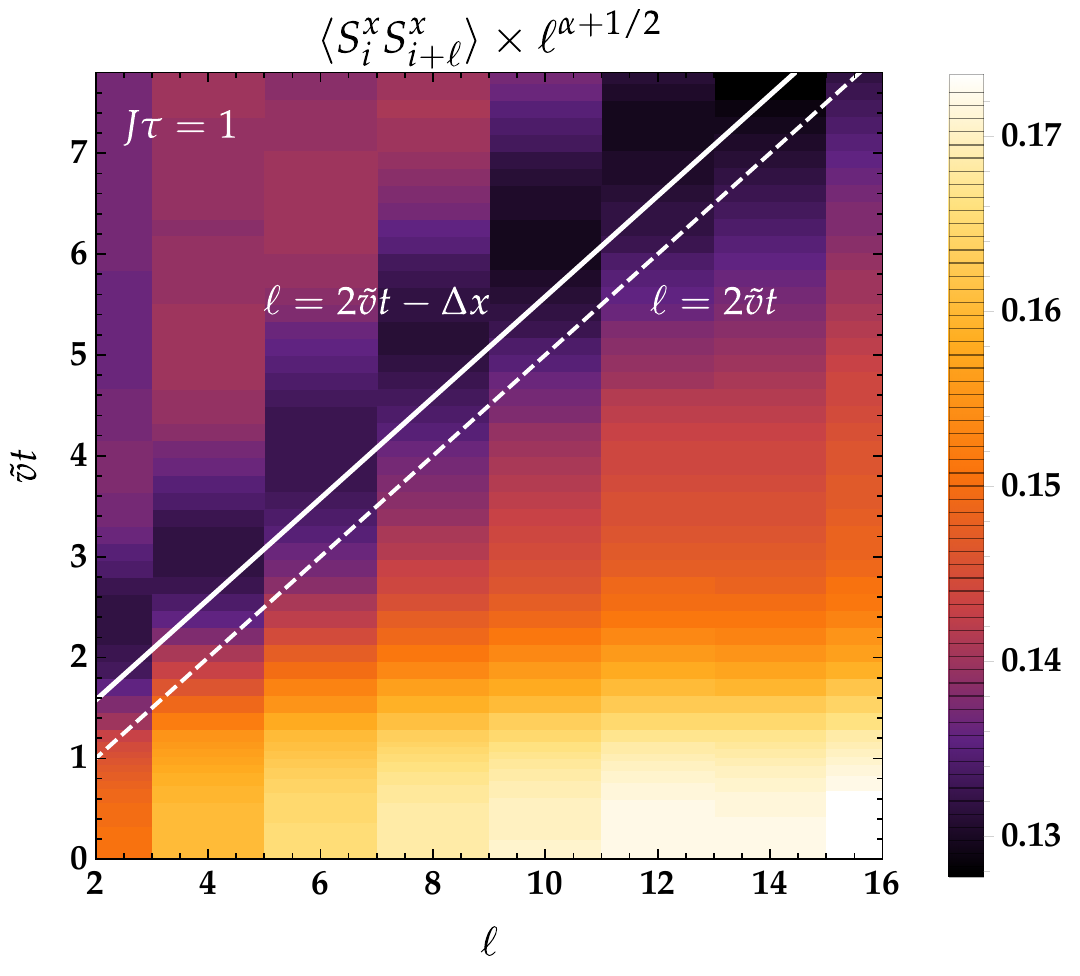}\\[4mm]
 \includegraphics[width=0.45\textwidth]{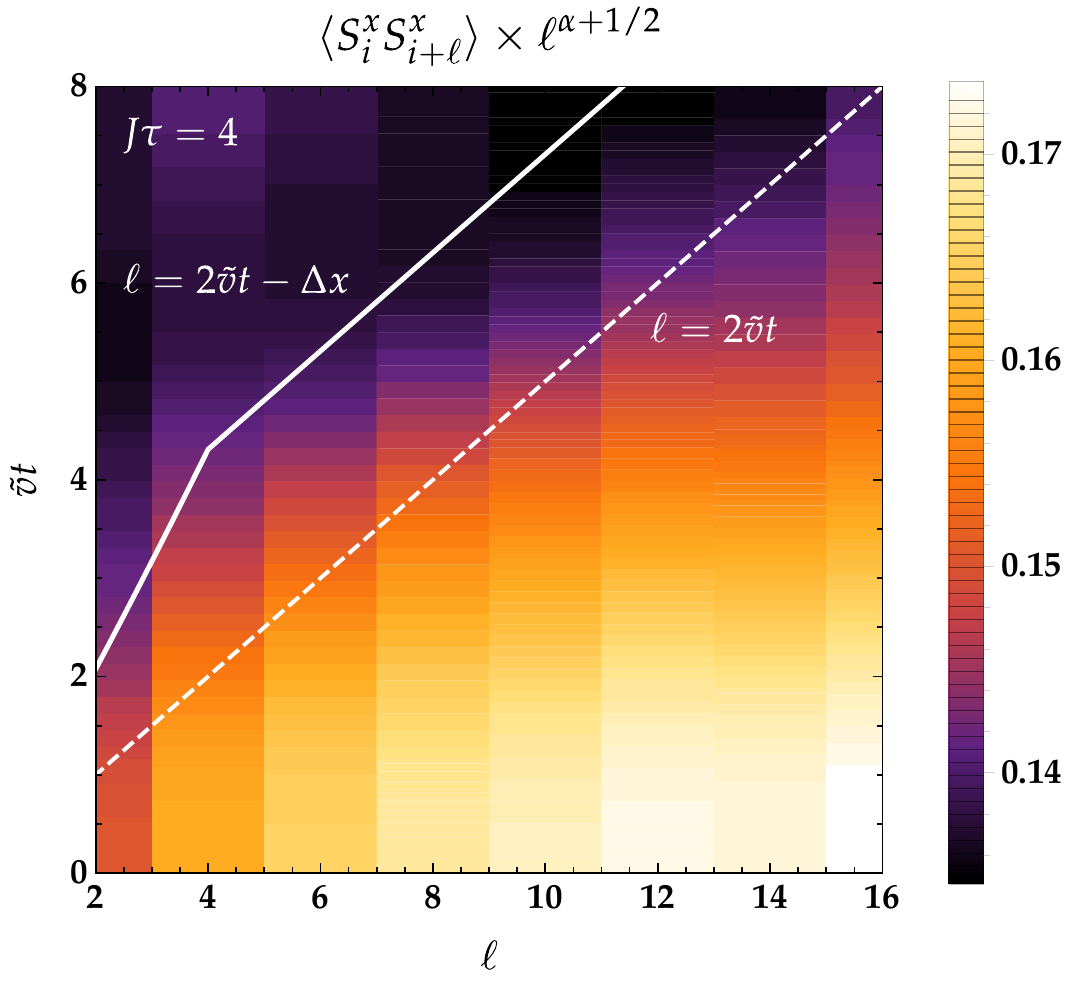}
\caption{(Colour online) Contour plot of the rescaled correlation function $\left\langle S^x_{i}(t) S^x_{i+\ell}(t)\right\rangle\ell^{\alpha+1/2}$ for a linear quench \eqref{eq:linear} with $\Delta=0.2$, $J\tau=1$ (upper panel) and $J\tau=4$ (lower panel). The exponent $\alpha$ used in the rescaling is defined in Eq.~\eqref{eq:alpha}. The solid white line indicates the position of the local minimum following the light cone $\ell\approx 2\tilde{v}t-\Delta x$, while the dashed line indicates the position of the light cone for a sudden quench of equal strength (ie, $\tau=\Delta x=0$).}
\label{fig:transversecontour}
\end{center}
\end{figure}
In this section we discuss the dynamics of the equal-time transverse two-point function \eqref{eq:transverse}. In Fig.~\ref{fig:transversecontour} we show contour plots of the correlation function in spacetime. We clearly observe a light cone, whose precise position is indicated by the solid white line together with the position of the light cone for a sudden quench of equal strength (dashed white line).  The light cone also shows a kink at $Jt=4$ in the lower panel, which originates from the end of the quench protocol at this time and thus a change in the effective velocity of the excitations. Before discussing the properties and origin of the observed light-cone behaviour in more detail, we compare our results to the field-theoretical predictions. 

\subsection{Field-theoretical prediction}
As is well established,\cite{Giamarchi04} the low-energy regime of the XXZ chain \eqref{eq:H} in equilibrium is described by the Luttinger model with the effective velocity and Luttinger parameter given by
\begin{equation}
  \tilde{v}=\frac{\pi J}{2}\frac{\sqrt{1-\Delta^2}}{\arccos \Delta},\quad
  K= \frac{\pi}{2}\frac{1}{\pi- \arccos \Delta}.
  \label{eq:LL}
\end{equation}
In this framework the leading contribution of the transverse correlation function at large separations is given by $\left\langle S^x_i S^x_{i+\ell} \right\rangle\simeq (-1)^\ell\,A_0^x\,\ell^{-1/(2K)}$. Here an exact analytical expression for the amplitude $A^x_0$ is known.\cite{LukyanovZamolodchikov97,Lukyanov99} 

As has been shown previously,\cite{Barmettler-10,KRSM12,Pollmann-13,Coira-13,Collura-15} various aspects of the quench dynamics of the XXZ chain can also be described using the Luttinger model. In particular, the time evolution of the two-point functions \eqref{eq:transverse} after a sudden quantum quench has been worked out with the result\cite{Cazalilla06,IucciCazalilla09,Collura-15}
\begin{equation}
\left\langle S^x_i (t) S^x_{i+\ell} (t) \right\rangle\simeq
(-1)^\ell \frac{A^x}{\sqrt{\ell}}\left\vert\frac{\ell^2-(2 \tilde{v} t)^2}{\ell^2\,(2\tilde{v}t)^2}\right\vert
^{\frac{\alpha}{2}}
\label{eq:SxSx}
\end{equation}
where
\begin{equation}
\alpha=\frac{1}{4}\left(\frac{1}{K^2}-1\right)
\label{eq:alpha}
\end{equation}
and the amplitude $A^x$ determined numerically.\cite{Collura-15} The analytical result \eqref{eq:SxSx} possesses a singularity at $\ell=2\tilde{v}t$, which physically originates in the light-cone effect to be discussed in Sec.~\ref{sec:lightcone} below.

Turning to finite-time quenches, the time evolution of two-point functions in the Luttinger model has been studied in several works.\cite{Dora-11,Pollmann-13,Bernier-14,CS16} In particular, the transverse correlation function \eqref{eq:transverse} after the quench, ie, at times $t>\tau$, takes the form\cite{CS16} 
\begin{equation}
\left\langle S^x_i (t) S^x_{i+\ell} (t)\right\rangle\simeq
(-1)^\ell \frac{\mathcal{A}^x}{\sqrt{\ell}}\left\vert\frac{\ell^2-(2 \tilde{v} t-\Delta x)^2}{\ell^2\,(2\tilde{v}t-\Delta x)^2}\right\vert
^{\frac{\alpha}{2}}.
\label{eq:SxSxwithlag}
\end{equation}
Again the correlation function possesses a singularity, but this time there is an additional lag $\Delta x$ in its position in space time. This lag $\Delta x$ can be attributed to the creation and propagation of quasiparticle pairs during the quench (see Sec.~\ref{sec:lightcone} below). The lag depends on the details of the quench protocol (see, eg, Fig.~\ref{fig:delay}), it is explicitly given by
\begin{equation}
\Delta x=\frac{4Kv_\text{F}\tau}{1-K^2}\left[\hat{g}_2(\tau)-\frac{1}{\tau}\int_{0}^{\tau}dt\,\hat{g}_2(t)\right],
\label{eq:lag}
\end{equation}
where, for the case of the XXZ chain \eqref{eq:H} we are considering here, the bare velocity and interaction parameter $\hat{g}_2$ are given by $v_\text{F}=J$ and
\begin{equation}
  \hat{g}_2(t) =2\sqrt{1-\Delta(t)^2}\frac{\pi+\arcsin\Delta(t)}{\pi^2 -4\arcsin^2\Delta(t)}\arcsin\Delta(t),
  \label{eq:g2}
\end{equation}
respectively. The result \eqref{eq:lag} was derived for short to moderate quench times $\tau$, which for the XXZ chain implies the region of applicability to be given by $J\tau\lesssim 1$. In addition, the amplitude $\mathcal{A}^x$ is expected to depend on the quench time and protocol. In the sudden-quench limit \eqref{eq:SxSxwithlag} reduces to \eqref{eq:SxSx} since $\Delta x\to 0$ and $\mathcal{A}^x\to A^x$. Finally, we note in passing that linear quenches \eqref{eq:linear} in the anisotropy do not translate into linear quenches in the corresponding interaction parameter $\hat{g}_2$.

\subsection{Simulation results}
\begin{figure}[t]
\begin{center}
 \includegraphics[width=0.45\textwidth]{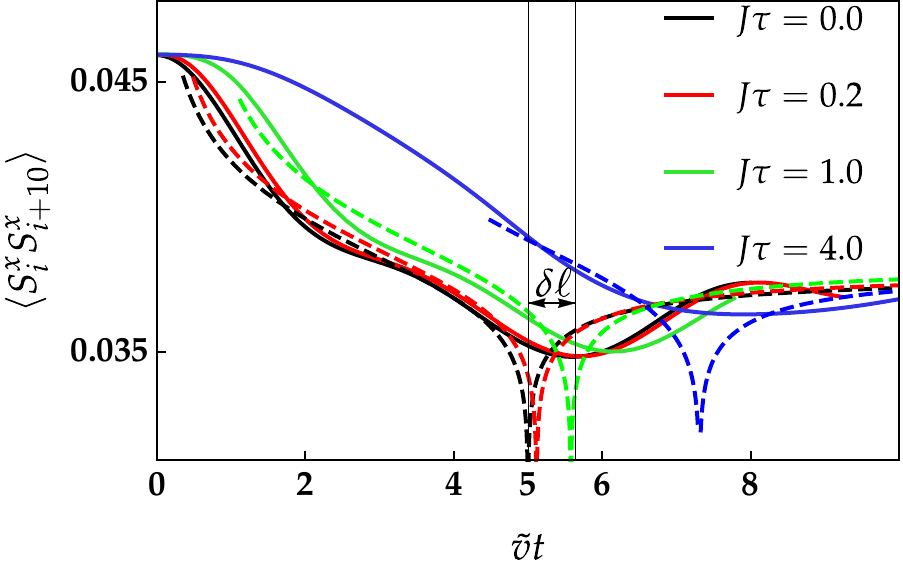}
\caption{(Colour online) Transverse two-point function \eqref{eq:transverse} for a distance of $\ell=10$ sites, a linear quench \eqref{eq:linear} with $\Delta=0.2$ and several quench times $\tau$. The dashed line shows the field-theoretical prediction \eqref{eq:SxSxwithlag} with the respective lags $\Delta x$. The amplitude is fitted in the sudden-quench case to $\mathcal{A}^x=0.1396$ and used for the other quench times as well. Even in the sudden-quench case we observe a delay $\delta\ell$ of the light-cone feature in the numerical data as compared to the analytical prediction, which originates from the finite bandwidth in the lattice model (see also Sec.~\ref{sec:lightcone}).}
\label{fig:cuts1}
\end{center}
\end{figure}
\begin{figure}[t]
\begin{center}
 \includegraphics[width=0.45\textwidth]{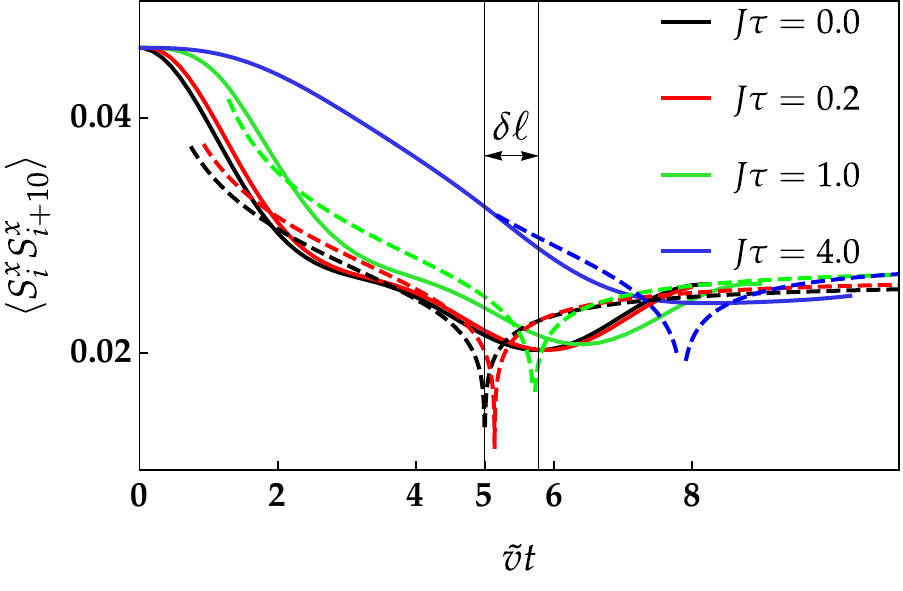}
\caption{(Colour online) Transverse two-point function as in Fig.~\ref{fig:cuts1} but for a linear quench \eqref{eq:linear} with $\Delta=0.5$. The amplitude of the analytical result has the value $\mathcal{A}^x=0.1287$.}
\label{fig:cuts2}
\end{center}
\end{figure}
\begin{figure}[t]
\begin{center}
 \includegraphics[width=0.45\textwidth]{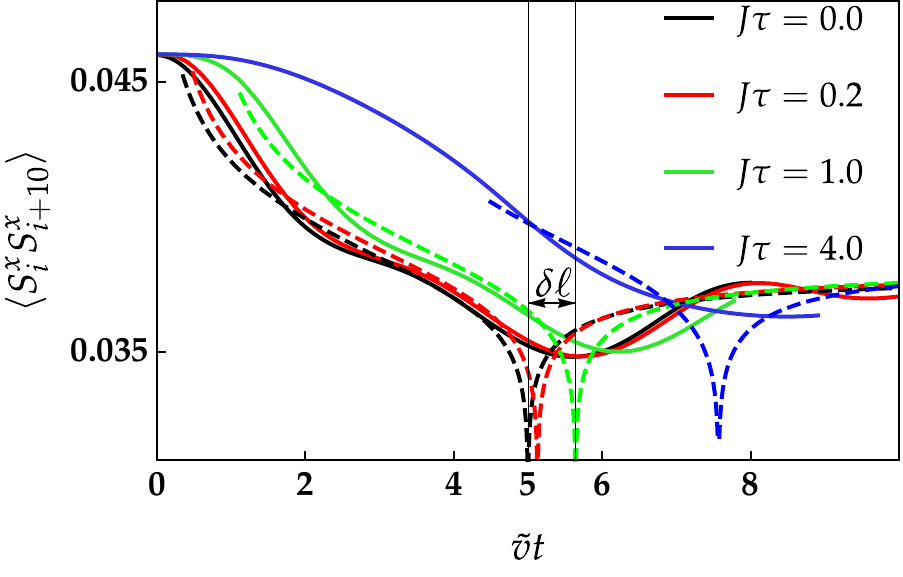}
\caption{(Colour online) Transverse two-point function as in Fig.~\ref{fig:cuts1} but for an exponential quench \eqref{eq:exponential} with $\Delta=0.2$. The amplitude again has the value $\mathcal{A}^x=0.1396$.}
\label{fig:cuts3}
\end{center}
\end{figure}
A more detailed understanding of the correlation function shown in Fig.~\ref{fig:transversecontour} can be obtained by considering cuts at finite separation $\ell$. Such cuts are shown in Figs.~\ref{fig:cuts1}--\ref{fig:cuts3} for $\ell=10$, for both linear and exponential quench protocols. We plot the numerical results (solid lines) together with the field-theoretical expression \eqref{eq:SxSxwithlag} including the lag \eqref{eq:lag} (dashed lines) for several quench times $\tau$. We have determined the amplitude $\mathcal{A}^x$ by fitting the prediction \eqref{eq:SxSxwithlag} for times inside the light cone, $2\tilde{v}t<\ell$. The resulting values in the sudden-quench case are in very good agreement with the ones obtained by Collura \emph{et al.}\cite{Collura-15} For finite-time quenches we observe a dependence of $\mathcal{A}^x$ on the precise fitting window, with the differences to the sudden-quench amplitude $A^x$ being less than $5\%$. Thus in the figures we use $\mathcal{A}^x=A^x$ for all quench times. As can be seen from the plots, the agreement between the numerical and analytical results is still very good, thus indicating that the dependence of the amplitude $\mathcal{A}^x$ on the quench duration is rather weak.

\begin{figure}[b]
\begin{center}
 \includegraphics[width=0.45\textwidth]{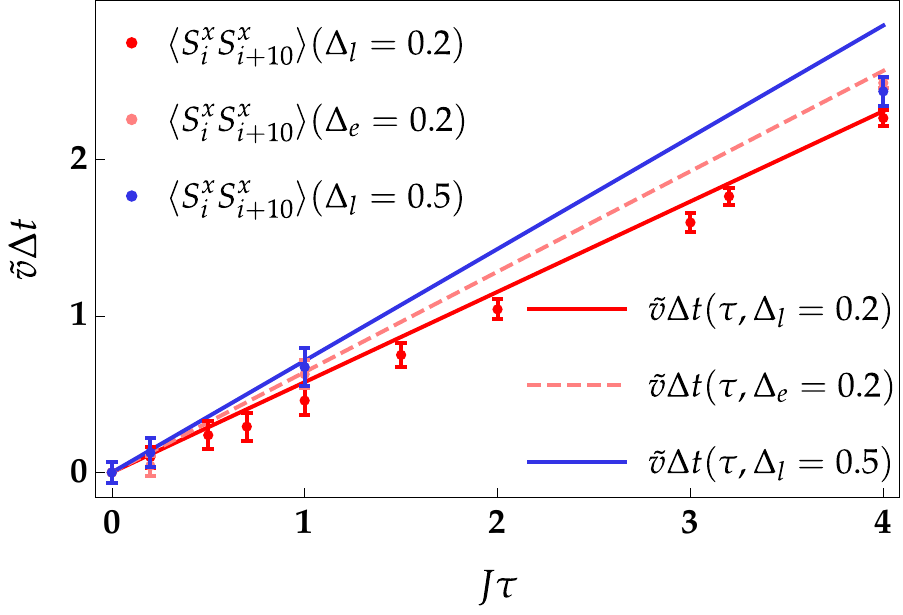}
\caption{(Colour online) Delay $\Delta t$ observed in the propagating minimum for a linear quench with $\Delta_l=0.2$ (red), exponential quench with $\Delta_e=0.2$ (pink) as well as linear quench with $\Delta_l=0.5$ (blue), as compared to the position of the minimum for the sudden quench (black line in Fig.~\ref{fig:cuts1}). The data points have been extracted from the numerical simulations, the lines show the prediction from the Luttinger model obtained using Eqs.~\eqref{eq:lag} and~\eqref{eq:g2} together with $\Delta x=2\tilde{v}\Delta t$. In particular, the delay depends on the details of the quench protocol, as can be seen from both the data and predictions for $\Delta_l=\Delta_e=0.2$.}
\label{fig:delay}
\end{center}
\end{figure}
Our simulation results clearly exhibit a minimum corresponding to the light cone already identified in the contour plot of Fig.~\ref{fig:transversecontour}. In comparison to the analytic result \eqref{eq:SxSxwithlag}, the singularity at $\ell=2\tilde{v}t-\Delta x$ is cut off by the finite bandwidth of the lattice model, thus turning the singularity into the propagating minimum. We also observe that the position of this minimum grows with the quench time $\tau$, in qualitative agreement with the result \eqref{eq:lag}. A more quantitative analysis is presented in the next section. Overall, we find good agreement between the field-theoretical result \eqref{eq:SxSxwithlag} and our numerical simulations, in accordance with the previously observed\cite{Collura-15} agreement for sudden quenches. 

\subsection{Light cone}\label{sec:lightcone}
\begin{figure}[t]
\begin{center}
 \includegraphics[width=0.45\textwidth]{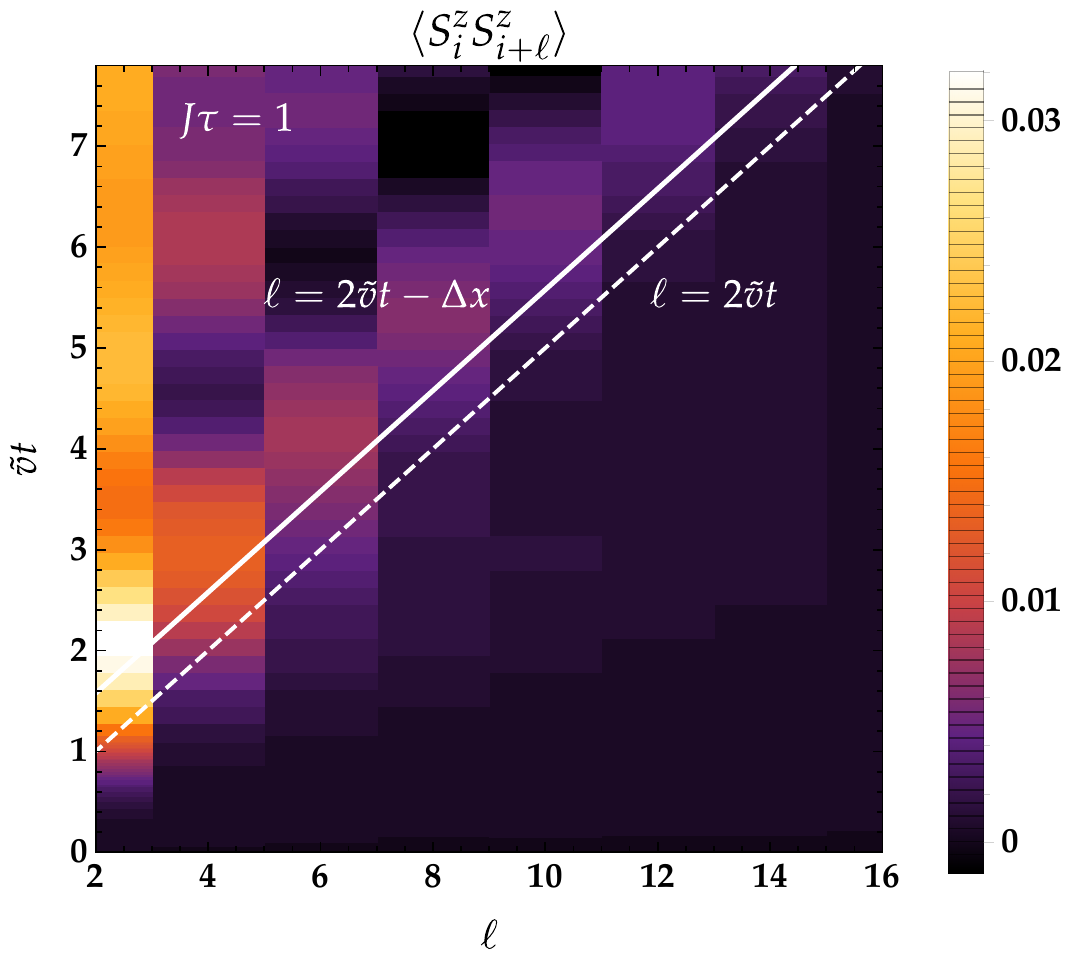}\\[4mm]
 \includegraphics[width=0.45\textwidth]{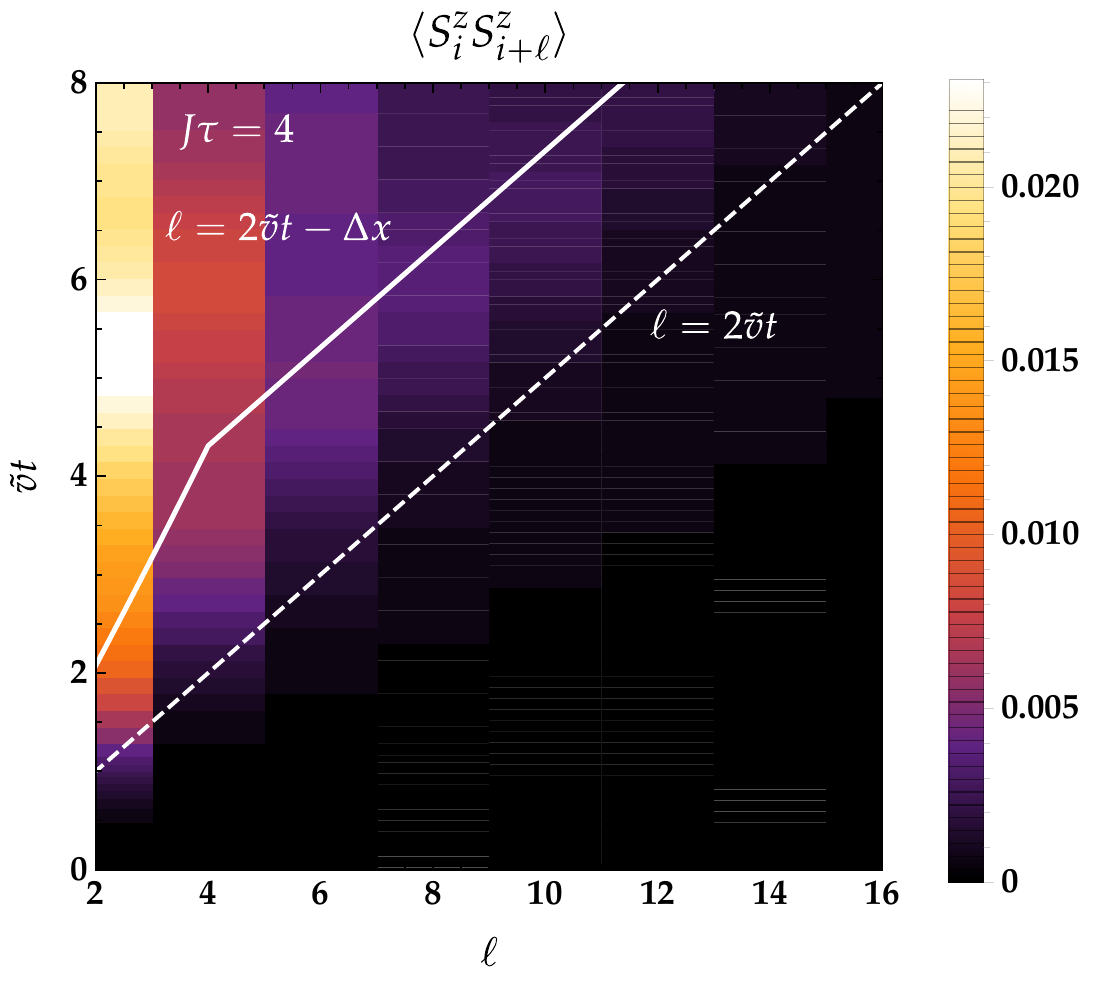}
\caption{(Colour online) Contour plot of the  correlation function $\left\langle S^z_{i}(t) S^z_{i+\ell}(t)\right\rangle$ for a linear quench \eqref{eq:linear} with $\Delta=0.2$, $J\tau=1$ (upper panel) and $J\tau=4$ (lower panel). The solid white line indicates the position of the local maximum following the light cone $\ell\approx 2\tilde{v}t-\Delta x$, while the dashed line indicates the position of the light cone for a sudden quench of equal strength (ie, $\tau=\Delta x=0$).}
\label{fig:longitudianalcontour}
\end{center}
\end{figure}
Let us now turn to a more detailed analysis of the light cone. As discussed above, the minimum in the  simulation data shown in Figs.~\ref{fig:cuts1}--\ref{fig:cuts3} corresponds to the singularity at $\ell=2\tilde{v}t-\Delta x$ in the analytical result \eqref{eq:SxSxwithlag}. However, even for the sudden quench, ie, the black solid and dashed lines, we see a delay $\delta\ell$ of the minimum as compared to the sharp singularity (see Fig.~\ref{fig:cuts1} for an illustration). This effect was also observed in previous works.\cite{Collura-15} It originates from cut-off effects, ie, the finite bandwidth in the lattice model \eqref{eq:H} not only leads to a softening of the singularity but also to a shift in its position to later times.\cite{footnote}

In order to analyse the additional delay $\Delta x$ caused by the existence of the finite quench time $\tau$, we subtract the delay $\delta\ell$ discussed above in the following. Thus in Fig.~\ref{fig:delay} we show the difference of the positions of the propagating minima for a finite-time quench and a sudden quench with identical final interaction strengths as a function of the quench time $\tau$. The shown error bars originate from the uncertainty in determining the minima in the time evolution. Specifically they are the combined imprecisions of the minima of the sudden and finite-time quenches due to the limited time resolution given by the step size of our simulations. We compare the delay $\Delta x$ thus extracted from the numerical simulations to the analytical prediction obtained in the Luttinger model, ie, Eqs.~\eqref{eq:lag} and~\eqref{eq:g2} together with $\Delta x=2\tilde{v}\Delta t$. We observe very good agreement between the two. Only for long quench times $J\tau\sim 4$ and stronger quenches ($\Delta=0.5$) quantitative deviations become visible. 

Furthermore, comparing the results for the linear and exponential quench with final anisotropy $\Delta=0.2$ we see that the in the latter case the delay $\Delta x$ is larger, again in agreement with the analytical prediction. Thus our numerical simulations support the qualitative picture that the details of the quench protocol influence the correlation functions and in particular the delay of the light cone even after the quench has ended.

Finally, let us comment on the physical origin of the light-cone delay: The quench excites entangled quasiparticles which propagate through the system at velocity $\tilde{v}$. For sudden quenches, all quasiparticles are excited at $t=0$, hence they induce correlations between two points $x_i$ and $x_j$ at time $t$ if $\vert x_i-x_j\vert<2\tilde{v}t$, ie, if the points can be reached by entangled quasiparticles.\cite{CalabreseCardy06,CalabreseCardy07} In contrast, for finite-time quenches we observe an additional delay $\Delta t$, ie, correlations appear only if $\vert x_i-x_j\vert<2\tilde{v}(t-\Delta t)$. There are two effects that physically explain this delay: (i) The excitation of quasiparticles will take place during the whole quench time $\tau$, ie, not just at $t=0$. (ii) The velocity of the quasiparticles during the quench is in general smaller than the the post-quench velocity $\tilde{v}$, thus for times $t<\tau$ quasiparticles will propagate at a slower velocity as compared to the sudden quench. Both effects\cite{Bernier-14,CS16} lead to a lag $\Delta x$ of the light cone as compared to the sudden quench, and thus to the delay $\Delta t$ shown in Fig.~\ref{fig:delay}. 
  
\section{Longitudinal two-point function}\label{sec:longitudinal}
\begin{figure}[t]
\begin{center}
 \includegraphics[width=0.45\textwidth]{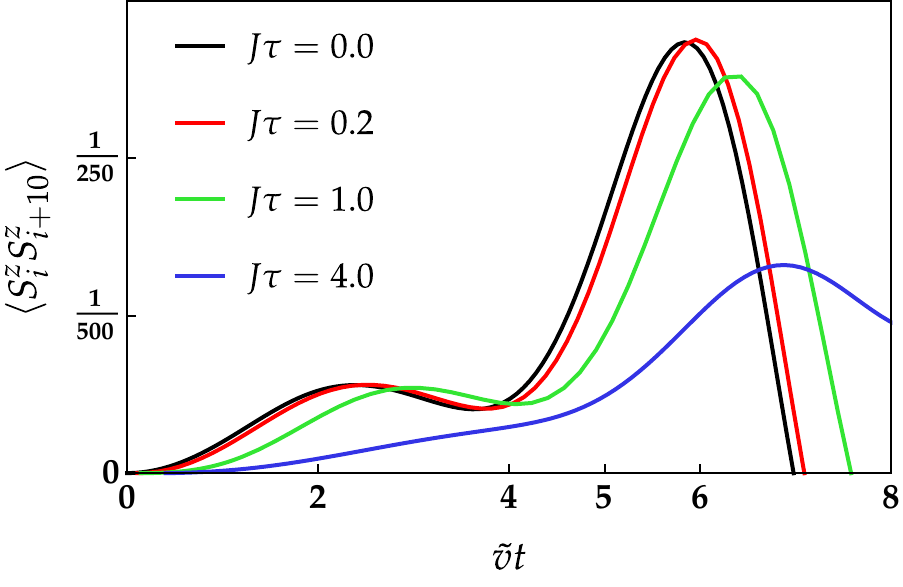}
\caption{(Colour online) Longitudinal two-point function \eqref{eq:longitudinal} for a distance of $\ell=10$ sites, a linear quench \eqref{eq:linear} with $\Delta=0.2$ and several quench times $\tau$.}
\label{fig:cuts4}
\end{center}
\end{figure}
\begin{figure}[t]
\begin{center}
 \includegraphics[width=0.45\textwidth]{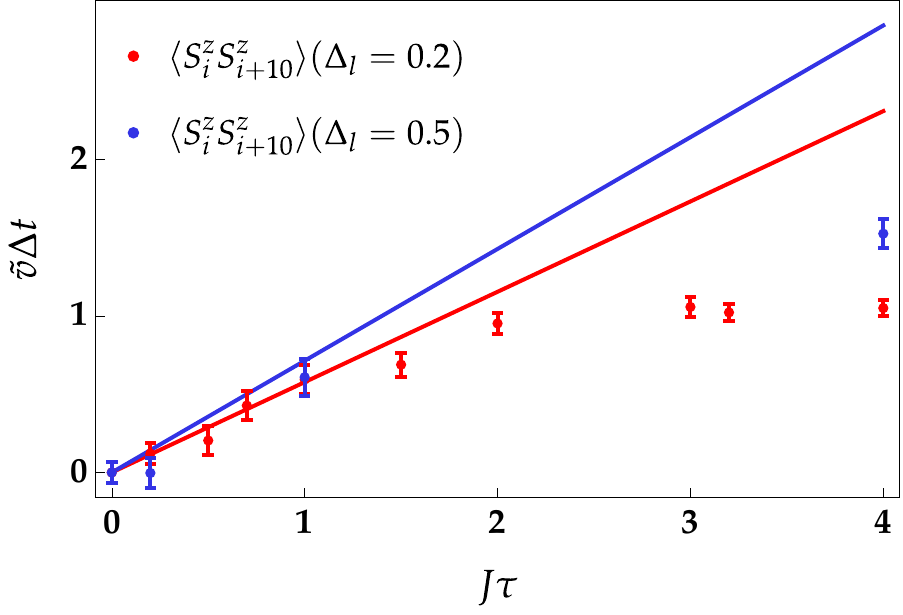}
\caption{(Colour online) Delay $\Delta t$ observed in the propagating maximum of the longitudinal correlation function for linear quenches to $\Delta=0.2$ and $\Delta=0.5$, as compared to the position of the maximum for the sudden quench (black line in Fig.~\ref{fig:cuts4}). For short to medium quench times, $J\tau\lesssim 2$, we observe good agreement between the numerical simulations and the prediction from the Luttinger model.}
\label{fig:delay2}
\end{center}
\end{figure}

Finally, let us briefly consider the longitudinal two-point function \eqref{eq:longitudinal}. Contour plots and  cuts are shown in Figs.~\ref{fig:longitudianalcontour} and~\ref{fig:cuts4}. As for the transverse correlation function above, we observe a clear delay of the light-cone feature for finite quench times $\tau$ as compared to the sudden-quench case. Further analysis of this delay (Fig.~\ref{fig:delay2}) shows again good agreement between the numerical results and the prediction \eqref{eq:lag}, thus confirming that the Luttinger model can also describe the delay of the light cone for the longitudinal correlation function.

\section{Conclusion}\label{sec:conclusions}
In this work, we have studied the time evolution of the correlation functions in the XXZ Heisenberg chain for interaction quenches of finite duration $\tau$. The quenches were performed in the critical regime, starting at the non-interacting point $\Delta=0$ and increasing the anisotropy either linearly or exponentially up to a final value $\Delta(t=\tau)>0$. We used a time-dependent DMRG algorithm to calculate the transverse and longitudinal equal-time spin-spin correlation functions at a distance $\ell$. Our results show a light-cone feature consistent with the quasiparticle picture originally put forward by Calabrese and Cardy.  \cite{CalabreseCardy06} Unlike the light cone resulting from a sudden quench, the light-cone front after finite-time quenches features a delay $\Delta t$. We have compared this delay with analytical results obtained for finite-time quenches in the Luttinger model.\cite{CS16} We find very good agreement for short and intermediate quench durations, with deviations only showing up for slow, strong quenches. We conclude that, despite the non-equilibrium nature of the quantum quench setup, the Luttinger model can be used to adequately describe features of the time evolution in the XXZ chain after a finite-time quantum quench. This is in particular true for the delay of the light cone caused by the finite quench duration.

\section{Acknowledgements}

We thank Mario Collura, Volker Meden, Neil Robinson and Peter Schmitteckert for very useful comments and discussions. This work is part of the D-ITP consortium, a program of the Netherlands Organisation for Scientific Research (NWO) that is funded by the Dutch Ministry of Education, Culture and Science (OCW). It was supported by the Netherlands Organisation for Scientific Research (NWO) under FOM 14PR3168.


\end{document}